# THE AI ASSESSMENT SCALE (AIAS) IN ACTION: A PILOT IMPLEMENTATION OF GENAI SUPPORTED ASSESSMENT-A PREPRINT

Leon Furze [1*], Mike Perkins [2], Jasper Roe [3], Jason MacVaugh [2]

[1] Deakin University, Australia
[2] British University Vietnam, Vietnam.
[3] James Cook University Singapore, Singapore.
[*] Corresponding Author: l.furze@deakin.edu.au

March, 2024

This is the preprint version of the paper 'The AI Assessment Scale (AIAS) In Action: A Pilot Implementation ff GenAI Supported Assessment', published in Australasian Journal of Educational Technology . Please refer to the published paper for the final text:

Furze, L., Perkins, M., Roe, J., & MacVaugh, J. (2024). The AI Assessment Scale (AIAS) in action: A pilot implementation of GenAI-supported assessment. *Australasian Journal of Educational Technology*. https://doi.org/10.14742/ajet.9434

## Abstract

The rapid adoption of Generative Artificial Intelligence (GenAI) technologies in higher education has raised concerns about academic integrity, assessment practices, and student learning. Banning or blocking GenAI tools has proven ineffective, and punitive approaches ignore the potential benefits of these technologies. This paper presents the findings of a pilot study conducted at British University Vietnam (BUV) exploring the implementation of the Artificial Intelligence Assessment Scale (AIAS), a flexible framework for incorporating GenAI into educational assessments. The AIAS consists of five levels, ranging from 'No AI' to 'Full AI', enabling educators to design assessments that focus on areas requiring human input and critical thinking.

Following the implementation of the AIAS, the pilot study results indicate a significant reduction in academic misconduct cases related to GenAI, a 5.9% increase in student attainment across the university, and a 33.3% increase in module passing rates. The AIAS facilitated a shift in pedagogical practices, with faculty members incorporating GenAI tools into their modules and students producing innovative multimodal submissions. The findings suggest that the AIAS can support the effective integration of GenAI in HE, promoting academic integrity while leveraging the technology's potential to enhance learning experiences.

## Introduction

The rapid advancement of Generative Artificial Intelligence (GenAI) technologies took the education sector by surprise, creating a wave of speculation and concern regarding the impact of the technology on academic integrity, assessment practices, and student learning (Roe & Perkins, 2023). While GenAI may offer potential benefits, such as personalised learning, writing assistance, and research capabilities (Chan & Hu, 2023), critics have highlighted risks, including copyright infringement, labour exploitation, environmental impact, bias, privacy, and the deskilling of both students and educators (Caplan, 2024; Selwyn, 2022, 2024).

Higher education institutions face the challenge of ensuring the fair and transparent use of GenAI while mitigating the risk of exacerbating existing inequalities, with low-income students and non-native English speakers of particular concern (Amano et al., 2023; Duah & McGivern, 2024). Current approaches to GenAI in higher education often focus narrowly on academic misconduct, limiting opportunities for students to engage with the technology in meaningful ways and ignoring broader ethical considerations (Cotton et al., 2023; Perkins, 2023; Plata et al., 2023; Uzun, 2023). Unfortunately, student perspectives have been largely absent from these discussions, despite students holding favourable views on using GenAI in their own learning (Chan & Hu, 2023; Chiu, 2024).

To address these challenges, we propose the Artificial Intelligence Assessment Scale (AIAS) as a flexible and adaptable framework for incorporating GenAI technologies into educational assessment (Perkins et al., 2023). This paper presents the findings of a pilot study conducted at BUV, exploring the effects of implementing the AIAS on academic misconduct, student achievement, and pedagogical practices.

The paper begins with a literature review examining the current state of GenAI in higher education, the limitations of existing approaches, and the need for a comprehensive framework. The five-point AIAS, ranging from "no AI" to "full AI", is then introduced, followed by a detailed case study of its implementation at British University Vietnam (BUV). The results of the pilot study are presented and discussed, highlighting the potential of the AIAS to support the ethical and effective integration of GenAI in educational assessment. Finally, the implications of this study for future research and practice are discussed.

## Literature

### Global responses to GenAI in HE

Initial reactions to GenAI in HE were marked by bans and restrictions, with institutions seeking to safeguard academic integrity (Cotton et al., 2023; Perkins, 2023b; Plata et al., 2023; Uzun, 2023). However, as the understanding of GenAI capabilities and ubiquity grew, institutions began to adopt more nuanced approaches (Fowler et al., 2023; Group of Eight Australia, 2023; Lodge et al., 2023; The Russell Group, 2024). Emerging themes in institutional policies still focus on academic integrity but also acknowledge the need to support student learning and understand the ethical considerations of these technologies. Globally, a similar pattern of outright bans, followed by more nuanced formal policies, has emerged. Across Asia, universities in Hong Kong (Cheung & Wong, 2023), Japan (Nagoya University, 2023), South Korea, Singapore, and India (Leung & Niazi, 2023) vacillated between banning GenAI and adapting their existing policies. Prominent universities in the US followed similar trajectories, ultimately revising their policies to allow for some use of GenAI (Harvard University, n.d.; Stanford University, 2023; Yale University, 2023).

### GenAI across academic disciplines

Studies have highlighted that GenAI has the potential to support students and enhance their learning experiences across various disciplines, including writing and composition (Cummings et al., 2024; Knowles, 2024), STEM fields (Amano et al., 2023; Cooper, 2023; Forero & Herrera-Suárez, 2023), creative disciplines (Bussell et al., 2023; Gozalo-Brizuela & Garrido-Merchan, 2023), and computer science (Liu et al., 2024). Like the broader policies, however, these studies have focused on the use of text-based models such as ChatGPT and have failed to address the growing capabilities of GenAI. As reflected in Yan et al.'s review of AI in education (2024), the focus across disciplines has been on a narrow range of uses, such as tutor-chatbots, low-stakes resource creation (e.g. quizzes), and error correction. Integrating GenAI effectively requires careful consideration of authentic assessment design, and the need to balance human and AI contributions (Miao & Holmes, 2023).

**Limitations of current approaches**

Despite the shift towards more inclusive stances on GenAI, current approaches often overemphasise academic misconduct, neglecting the potential benefits of the technologies (Birks & Clare, 2023; Luo, 2024). Knight et al.'s (2023) analysis of submissions to the Australian parliamentary inquiry into GenAI in education found a common theme in the need to prepare students to use GenAI; however, this has not yet been evidenced in the practical application of HE policies. Although included in various policies, sufficient student support is not yet evident, and students report feeling unprepared to use GenAI (HEPI, 2024; Kelly et al., 2023). One issue in supporting students to use the tools is the limited scope in understanding the capabilities of GenAI beyond text-based applications, with policies and student guidelines, including those in prominent associations like the Australian 'Group of Eight', treating GenAI as metonymical with ChatGPT (Australian National University, 2023a, 2023b; Monash University, 2023; University of Adelaide, 2023; University of Melbourne, 2023a, 2023b; UNSW, 2024). To ensure the authenticity of student work, some policies and guidance for educators suggest methods for "AI-proofing" assessments, including analysis of videos and images, or use of recorded videos of students (Duah & McGivern, 2024; Rudolph et al., 2023). However, given recent advances in GenAI technology, which include image recognition (OpenAI, 2023), the ability to interpret video (Pichai & Hassabis, 2024), video generation (OpenAI, 2024) and 3D, VR, and AR (Bussell et al., 2023; Gozalo-Brizuela & Garrido-Merchan, 2023), these suggestions may be short-lived.

**Ethical Implications of GenAI tools in Education**

The integration of GenAI into education raises a multitude of ethical concerns that must be addressed to ensure responsible adoption and mitigate potential risks. One of the primary issues in GenAI systems is bias and fairness. The datasets used to train these models, such as ImageNet and The Pile, often reflect the biases and worldviews of the predominantly white, middle class, and English-speaking males who contribute to them (Bender et al., 2021; Crawford, 2021). These biases can manifest in the output of GenAI tools, leading to discriminatory or offensive content (Sun et al., 2023). Attempts to filter such content can inadvertently result in further bias; for example, by flagging words with religious connotations or those more frequently used in LGBTI communities (EFRA, 2022).

Privacy and data security are also significant concerns in the context of GenAI in education. The use of AI technologies, including both predictive and generative AI, can contribute to the "datafication" of students, where data is collected from various aspects of their lives, often for profit (Eynon, 2022; Pangrazio & Sefton-Green, 2022). Intellectual property rights and copyright infringement are also areas of concern for GenAI. The use of copyrighted material in the datasets used to train GenAI models may constitute an unethical infringement of intellectual property rights (Perrotta et al., 2022). Although legal approaches vary internationally (Ozcan et al., 2023), this issue remains unresolved, with ongoing lawsuits related to the inclusion of copyrighted material in the GenAI tool datasets.

Addressing these ethical implications is not only a matter of responsible technology integration but also an educational imperative. As GenAI becomes increasingly prevalent in society, educators have an ethical duty to prepare students for an AI-focused world postgraduation. By prohibiting the use of GenAI tools or labelling their use as plagiarism and misconduct, educators may be doing a disservice to their students by failing to equip them with the skills needed for the future (Anson, 2022). The view that GenAI tools fundamentally threaten academic integrity and enable plagiarism is an oversimplification of this complex technology, which lacks nuance. Instead, institutions must develop policies that promote the ethical and transparent use of GenAI considering the multifaceted nature of these technologies and their potential to enhance learning experiences when used responsibly.

## The AI Assessment Scale (AIAS)

We propose the AI Assessment Scale (AIAS) to promote the transparency and ethical use of GenAI tools. The AIAS is designed to be flexible, clear for both educators and students with limited knowledge of new AI technologies, and adaptable across a wide range of disciplines and contexts. The following is a summary of the five levels of the AIAS, with brief examples. The AIAS has been discussed in detail by Perkins et al. (2023).

### Level 1 - No AI

Students complete assessments without any use of GenAI tools. This may be for practical reasons, for example, a practical task with no electronic devices which precludes the use of GenAI or for assessment security purposes, for example, invigilated, technology-free examinations. Tasks at this level may be discussions, debates, or technology-free group activities, where the use of GenAI tools would either not be beneficial to the students' learning or would disguise whether learning objectives were being met.

### Level 2 - AI-Assisted Idea Generation and Structuring

AI is used for brainstorming and working with ideas or notes; however, the final submission must be free of any GenAI content. The task may permit, for example, the use of AI-assisted Automatic Speech Recognition (ASR) to transcribe notes, the use of GenAI to convert notes into outlines, or to contribute to brainstorming or suggestions for improvement on already created work. Image generation technologies might be used to generate starting points for designs in art- or design-based subjects or tools used to explore possibilities to produce software in computer science subjects, but no GenAI-created content can be included in the final submissions.

### Level 3 - AI-Assisted Editing

AI can be used to edit student-generated work; however, the original work must be provided for comparison. This level permits the use of tools which can support rewriting and editing to clarify ideas created by students, the use of GenAI for editorial purposes, or editing text captured with ASR (e.g. verbally recorded drafts). In a multimodal context, AI-assisted editing tools could be permitted alongside documentation of the process.

### Level 4 - AI Task Completion, Human Evaluation

AI is used to complete major elements of the task, with students critiquing and reflecting on AI-generated content. At this level, students might create significant portions of the outcome with AI, and then reflect on the quality, voracity, bias, or overall quality of the AI-generated data. For example, AI may be used to create mock datasets in the sciences, entire written responses to literature, or complete code. The core element of this level is that students are required to reflect on and assess these GenAI-created outputs, and not just to use them to complete a set task.

### Level 5 - Full AI

AI is used throughout the assessment without the need to specifically acknowledge any AI-generated content. At this level, the use of any multimodal GenAI technology is either permitted for the completion of the task or required to be used to score highly. For example, requiring students to use AI avatars in the production of video content, or to write responses using GenAI text editing tools alongside their written work. This level is particularly suitable when one of the learning objectives is related to the use of GenAI, but it may be integrated into any assessment with the recognition that students will be expected to use these tools alongside their own work in future work environments.

The five-point AIAS is presented in Table 1.

| 1 | NO AI | The assessment is completed entirely without AI assistance. This level ensures that students rely solely on their knowledge, understanding, and skills.<br><br>**AI must not be used at any point during the assessment.** |
|---|---|---|
| 2 | AI-ASSISTED IDEA GENERATION AND STRUCTURING | AI can be used in the assessment for brainstorming, creating structures, and generating ideas for improving work.<br><br>**No AI content is allowed in the final submission.** |
| 3 | AI-ASSISTED EDITING | AI can be used to make improvements to the clarity or quality of student created work to improve the final output, but no new content can be created using AI.<br><br>**AI can be used, but your original work with no AI content must be provided in an appendix.** |
| 4 | AI TASK COMPLETION, HUMAN EVALUATION | AI is used to complete certain elements of the task, with students providing discussion or commentary on the AI-generated content. This level requires critical engagement with AI generated content and evaluating its output.<br><br>**You will use AI to complete specified tasks in your assessment. Any AI created content must be cited.** |
| 5 | FULL AI | AI should be used as a "co-pilot" in order to meet the requirements of the assessment, allowing for a collaborative approach with AI and enhancing creativity.<br><br>**You may use AI throughout your assessment to support your own work and do not have to specify which content is AI generated.** |

*Table 1.* The AI Assessment Scale

# Case study: The AIAS in British University Vietnam

**Background**

British University Vietnam (BUV) is a private institution with approximately 2500 students and follows a UK-based curriculum with international accreditation through the UK Quality Assurance Agency (QAA). The language of instruction is English however, most students are non-native English speakers. In early 2023, discussions between the authors led to the adaptation of an initial AI assessment scale (Furze, 2023) into its current format, designed to support educators and students at BUV in developing assessment tasks and discussing the appropriate use of GenAI.

**GenAI policy adjustments**

BUV's reaction to GenAI tools was shaped by a gradual increase in the institutional knowledge of this disruptive technology. Initially, the technology went unnoticed, but as the number of suspected GenAI-related submissions and academic misconduct cases increased, BUV recognised the need for policy changes. In March 2023, students were informed that under the existing assessment regulations, the use of GenAI tools was not allowed in written work. This ban coincided with the launch of Turnitin's AI Detect feature, and faculty members were trained in the use of the detection tool. This also corresponded with the release of the more capable GPT-4 model from OpenAI, meaning that student use of ChatGPT would be harder to detect (Perkins, Roe, et al., 2023) produced by this model was more challenging to detect.

By July 2023, approximately 70% of all reported violations of academic integrity were AI related. Numbering in the hundreds, the penalties applied to these cases drew the attention of the Dean and Deputy Vice-Chancellor. With a consideration of the global news storm on the pervasiveness of AI usage in society (Roe & Perkins, 2023), BUV was faced with a dilemma familiar to HE institutions worldwide: maintaining academic integrity in the face of GenAI while still adequately preparing students for the use of these tools in their future careers. Research conducted at BUV also revealed limitations in AI detection tools (Perkins & Roe, 2023), reinforcing the need for a more nuanced approach to GenAI in assessments was required.

**Launch and implementation of the AIAS**

Considering these factors, in August 2023, BUV leadership decided to accept the use of GenAI tools for some student submissions. They set up a team to incorporate AI into teaching, learning, and assessment regulations with the following objectives:

1. Help educators consider how their assessments may need to be adjusted considering GenAI tools.
2. Clarify to students how and where GenAI tools might be used in their work.
3. Support students in completing assessments in line with the principles of academic integrity, thereby reducing the number of academic misconduct violations within BUV.

The scale was designed during August 2023 and was then approved by the Learning and Teaching Committee and the Academic Board of the University Senate in September 2023, accompanied by policy documentation explaining acceptable GenAI use. This policy focused on three core areas: Ethics & Transparency, Security & Privacy, and Limitations & Bias, with all training materials related to the AIAS being framed with these principles in mind. Policy documentation, training material, and the AIAS were implemented in October 2023 for the start of the new semester.

BUV faculty were briefed on the AIAS and trained on its application in assessments during the implementation period. The Vice Chancellor's decision to embrace GenAI tools required significant changes in the overall assessment strategy, including frequent opportunities for students to engage with GenAI in a positive and ethical manner.

Academic staff members were required to restrict Level 1 use to examinations, assessments set by professional bodies, or activities requiring live demonstrations of competence. Level 2 was primarily restricted to English language programs or required approval from Discipline Leaders. Setting assessments at Levels 3, 4, and 5 was encouraged to maximise learning opportunities while reducing instances of academic misconduct.

## Results

Following the October teaching semester and subsequent assessment period, the implementation of the AIAS showed initial signs of success in several key areas.

**Academic misconduct**

Data from the October 2023 semester shows the positive impact of the AIAS on academic integrity compared to prior semesters. In the April 2023 semester, 2.84% of the student submissions (116) received a GenAI-related academic misconduct penalty. Following the conclusion of the October 2023 semester, there were no cases of AI related academic misconduct. The zero cases of AI-related academic misconduct reported following the implementation of the AIAS may be viewed with cautious optimism. Although it appears that the AIAS scale has significantly decreased academic misconduct, it is important to factor in recent policy changes that permit wider use of General AI (GenAI) and give academic staff more freedom to modify grades. For instance, if the use of GenAI tools is permitted during an assessment but is exploited, such as using GenAI to generate the final text in a level three assessment where the student's original work does not show that the ideas belong to the student, faculty members can now exercise their academic judgment and reduce the marks accordingly, without having to report the case as misconduct. This change in policy may contribute to the decrease in reported cases of academic misconduct, as actions that were previously considered as misconduct can now be addressed through grade adjustments at the discretion of the academic staff.

**Improvements in student performance and engagement**

Looking at the average grades for the semester and comparing them with those from the same semester in the previous year may provide additional insights into the effectiveness of these policy changes. Reviewing the average grades achieved by students across all years and modules showed an average increase of 5.9% between the September 2022 semester (the first semester in which GenAI tools had gained significant media attention) and the September 2023 semester. This rise in academic performance was accompanied by a 33.3% increase in overall module pass rates. Although improvements in module averages are modest, this could imply that because of the introduction of the AIAS and the subsequent normalisation of GenAI tools, students are harnessing these more effectively for their studies. The dramatic increase in pass rates suggests that students for whom language skills may have been an impediment to expressing their ideas are now able to do so more effectively. Combined, these increases hint at these tools potentially aiding students in overcoming language barriers, suggesting that GenAI tools, when used within the framework of the AIAS, could enhance their overall educational experience.

**Changes in pedagogical practices**

The impact of the AIAS also needs to be considered within the broader context of the changes made in pedagogical practices related to GenAI at BUV. Discipline Leaders have reported positive shifts in pedagogical practices within their subjects, with many academic staff members choosing to incorporate GenAI tools into their modules. This aligns with the first aim of the AIAS intervention, which seeks to encourage educators to consider how assessments might be adjusted by considering GenAI capabilities. The innovative assessment designs that have emerged, particularly those that use GenAI as a "co-pilot", indicate a significant pedagogical shift towards embracing these technologies within the academic setting. The pilot implementation also revealed instances of highly creative uses of GenAI in student submissions. This has been particularly notable at the higher end of the AI scale, where students have engaged with a broad spectrum of GenAI tools to create complex and sophisticated multimodal submissions.

These projects have shown an increase in student engagement, particularly in areas where the traditional emphasis on long-form writing may have been a hurdle for second-language speakers, and AI tools have offered alternative means for students to express their knowledge and creativity. For instance, in courses where a module assessment may have been 100% based on an individual essay, providing students with the option to demonstrate their achievement of the learning outcomes in other formats, such as presentations aided by AI-generated visuals or summarising complex ideas with the help of AI, can help to level the playing field for those for whom English is not their first language. These outcomes suggest that the AIAS framework not only addresses concerns about academic integrity but also creates new avenues for student creativity and engagement.

Although the full impact of this initiative will become clearer as we continue to observe its long-term effects on academic standards and student learning outcomes, the initial introduction has demonstrated the potential for the AIAS to have a significant impact on supporting the ethical use of GenAI in assessments, while at the same time increasing engagement in new technology from both academic staff and students.

## Discussion

The AIAS aims to address the issues highlighted in the introduction, including supporting student learning, acknowledging the growing complexity of GenAI-based tools, and supporting the ethical and transparent use of these tools in education. It sets clear expectations for academic integrity and conveys these guidelines through an accessible five-point scale. By having a clear, flexible, and adaptable tool, the AIAS coherently addresses the current lack of a robust policy framework on GenAI use in higher education (Lodge et al., 2023) and enables assessments which emphasise authentic engagement with AI (Lodge et al., 2023). The AIAS is designed to embrace the full spectrum of GenAI capabilities, including text, image, audio, video, and code generation. By equipping students to engage with advanced GenAI technologies, the AIAS empowers educators to shift their focus back to the human aspects of learning (Miao & Holmes, 2023) and to design assessments that are primarily meaningful for learning (Luo, 2024).

**Generative AI, academic integrity, and assessment**

To encourage practices in education which acknowledge the potential of GenAI and support students in all disciplines to use these technologies appropriately, the narrative surrounding GenAI in HE must shift beyond "cheating". Concerns about academic dishonesty are not new, and misconduct in assessment tasks using AI can be seen as an extension of preexisting student behaviours (Birks & Clare, 2023). This shift in narrative is particularly important in the context of this case, as Vietnamese EAP learners have traditionally been perceived as passive and autonomous (Roe & Perkins, 2020). However, research has shown that these misconceptions do not accurately reflect students' willingness to engage in autonomous learning practices when provided with necessary support, guidance, and tools.

As with pre-AI discussions of academic integrity, it is important to recognise that students' misconduct is not equal across all assessment types; some forms lend themselves to academic dishonesty more than others (Bretag et al., 2019). Certain assessment types, particularly those conducted online and without supervision, are likely to create more opportunities for misconduct (Roe et al., 2023). As such, thoughtful assessment and curriculum design can reduce the temptation for students to commit academic misconduct (Sutherland-Smith & Dawson, 2022), while clearly articulating guidelines and regulations can reduce opportunities for AI-related academic dishonesty in ways which appeal to individuals' consciences (Birks & Clare, 2023). The AIAS, with its structured, yet adaptable framework, could serve as a mechanism for this cultural shift. It not only delineates clear boundaries and expectations for GenAI use in academic tasks, but also encourages educators and students alike to explore the multifaceted capabilities of these technologies within an ethical framework. This contributes to transparent communication and collaboration, which has been identified as one of the most important points in developing an effective, trusting, and caring academic culture (Luo, 2024).

It is also necessary to critique the emergent proliferation of GenAI detection tools and their roles in academic integrity conversations. Given the pace of change in technology and methods which can be used to produce more sophisticated and human-like texts, the "arms race" (Cole & Kiss, 2000; Roe & Perkins, 2022) between GenAI tools and AI text detection software is likely to continue for some time. Although claims by software providers such as Turnitin and GPTZero that their software is accurate in detecting AI-generated content in student work, a growing body of empirical research has shown that these claims are not entirely accurate and that detection tools can be easily evaded (Anderson et al., 2023; Chaka, 2023; Chakraborty et al., 2023; Elkhatat et al., 2023; Perkins, Roe, et al., 2023; Weber-Wulff et al., 2023), with additional concerns regarding potential false positives among non-native English speakers (Liang et al., 2023).

Finally, although the training of staff to identify GenAI-created texts is somewhat effective (Abd-Elaal et al., 2022), the continuing development of GenAI tools may render these approaches ineffective. The ongoing tension between advancements in GenAI and detection methodologies underscores the potential limitations of relying solely on technology to ensure academic integrity. This highlights the importance of developing robust educational strategies, such as the AIAS, that emphasise ethical use, critical engagement, and creative incorporation of GenAI in learning and assessment, rather than focusing predominantly on detection and deterrence.

## Limitations

The initial findings from the AIAS pilot study offer promising insights into the potential integration of GenAI into HE assessment; however, these results are preliminary, and have several limitations. The evidence base for the effectiveness of the AIAS is still in its infancy, and the pilot's scope within a single institution, may not fully capture the scale's applicability across a range of educational contexts and disciplines. The positive outcomes reported require further validation through larger, more diverse studies to understand their generalisability. It is also necessary to acknowledge the AIAS has only been tested in a transnational educational environment rooted in a traditional Western academic cultural paradigm. Other forms of knowledge assessment and knowing may not translate well into the AIAS, representing an area for further study across different cultural contexts, pedagogies, and educational spaces.

The rapid development of GenAI technologies poses a challenge to the long-term relevance and applicability of the AIAS. The current iteration of the scale may not fully account for future advancements in GenAI capabilities, which could introduce new ethical, pedagogical, and assessment-related challenges not considered in the pilot study.

However, the observed changes in pedagogical practices and increased student engagement with GenAI tools following the implementation of the AIAS suggest that the framework has the potential to foster a more inclusive and adaptive learning environment. Finally, the reduction in reported cases of academic misconduct following the implementation of the AIAS might be influenced by factors other than the scale's introduction, and without a control group or comparative analysis, attributing the decrease solely to the AIAS's influence remains speculative.

## Conclusion

The pilot implementation of the AIAS at BUV demonstrates a pragmatic approach to integrating GenAI into assessment strategies within an HE context in an ethical, transparent, and effective manner. The AIAS framework offers a structured yet flexible approach that can adapt to the diverse needs of different disciplines and assessment types, providing clear guidelines for both students and educators regarding the appropriate use of GenAI tools. The positive outcomes observed in BUV, including a decrease in academic misconduct and improved student engagement and performance, suggest that when GenAI is used within well-defined ethical parameters, it can significantly enhance educational experiences and outcomes.

Future research should focus on expanding the evidence base for the effectiveness of the AIAS, exploring its applicability across different educational contexts, and refining the framework in response to the evolving capabilities of GenAI technologies. This will require a collaborative effort among educators, policymakers, and researchers to ensure that the integration of GenAI into higher education is ethical, equitable, and enhances the learning experience for all students. The AIAS offers a model for how institutions can address the challenges of academic integrity in the age of AI while leveraging technology to create more engaging and inclusive learning environments. As GenAI becomes an integral part of the professional and personal landscape, the AIAS can help higher education institutions prepare students for success in an increasingly AI-driven world.

### AI Usage Disclaimer

 This study used Generative AI tools for revision and editorial purposes throughout the production of the manuscript. Models used were ChatGPT (GPT-4) and Claude 3 (Opus). The authors reviewed, edited, and take responsibility for all outputs of the tools used in this study.